# UNIVERSAL SOFTWARE PLATFORM FOR VISUALIZING CLASS *F* CURVES, LOG-AESTHETIC CURVES AND DEVELOPMENT OF APPLIED CAD SYSTEMS

*Rushan Ziatdinov*[*]
*Ph.D. in Mathematical Modeling, Numeric Methods, Program Software*
*Department of Industrial and Management Engineering,*
*Keimyung University, Daegu, Republic of Korea*
*E-mail: ziatdinov@kmu.ac.kr, ziatdinov.rushan@gmail.com*
*URL: http://www.ziatdinov-lab.com/*
*ORCID: 0000-0002-3822-4275*

*Valerijan G. Muftejev*
*Ph.D. in Technical Sciences*
*Department of the Fundamentals of Mechanisms and Machines Design,*
*Ufa State Aviation Technical University,*
*Ufa, Russian Federation*
*E-mail: muftejev@mail.ru*
*URL: http://spliner.ru/*
*ORCID: 0000-0003-4352-3381*

*Rustam I. Akhmetshin*
*Senior Lecturer*
*Department of the Fundamentals of Mechanisms and Machines Design,*
*Ufa State Aviation Technical University,*
*Ufa, Russian Federation*
*E-mail: ahmetshin@bk.ru*
*ORCID: 0000-0003-2448-0654*

*Alexander P. Zelev*
*Ph.D. in Technical Sciences*
*Department of the Fundamentals of Mechanisms and Machines Design,*
*Ufa State Aviation Technical University,*
*Ufa, Russian Federation*
*E-mail: zelev7@mail.ru*
*ORCID: 0000-0002-1999-9729*

*Rifkat I. Nabiyev*
*Ph.D. in Education*
*Department of the Fundamentals of Mechanisms and Machines Design,*
*Ufa State Aviation Technical University,*
*Ufa, Russian Federation*
*E-mail: dizain55@yandex.ru*
*ORCID: 0000-0002-0920-6780*

---

[*] Corresponding author. Postal address: Department of Industrial and Management Engineering (Office 4308), College of Engineering, Keimyung University, 1000 Shindang-dong, Dalseo-Gu, 704-701 Daegu, Republic of Korea.

*Albert R. Mardanov*
*Engineer*
*Department of the Fundamentals of Mechanisms and Machines Design,*
*Ufa State Aviation Technical University,*
*Ufa, Russian Federation*
*E-mail: systems@list.ru*
*ORCID: 0000-0002-6635-8887*

**Abstract:** This article describes the capabilities of a universal software platform for visualizing class *F* curves and developing specialized applications for CAD systems based on Microsoft Excel VBA, the software complex FairCurveModeler, and computer algebra systems. Additionally, it demonstrates the use of a software platform for visualizing functional and log-aesthetic curves integrated with CAD Fusion360. The value of the curves is evident in visualizing the qualitative geometry of the product shape in industrial design. Moreover, the requirements for the characteristics of class *F* curves are emphasized to form a visual purity of shape in industrial design and to provide a positive emotional perception of the visual image of the product by a person.

> "There are no barriers to human thought."
>
> *Sergei Korolev*

**1. Introduction**

There is an extensive class of products with functional curves and surfaces that determine their essential design characteristics. These are, for example, external contours of aircraft, vessels, working surfaces of soil-cultivating units, cam profiles, and axial lines of the road route, airfoils, turbine blades, and compressors.

In [1-2], basic requirements to the quality of functional curves are constructed by motion analysis of a material point along the trajectory of a spatial curve. These requirements, as generalized in [3], relate to the smoothness criteria and are invariant with respect to the specific operation of a particular product. They are as follows:

1. The order of smoothness not lower than 4
2. Absence or the minimal number of curvature extremes
3. Small values of variation for the curvature and the rate of its change
4. The small value of the potential energy of the curve
5. Aesthetic analysis from the standpoint of the laws of technical aesthetics.

Curves of high quality that satisfy the above smoothness criteria are called *class F curves* (Functional) [3-5], in contrast to the term *class A* (Aussenhaut[1]) *curves*, which was borrowed by Gerald Farin from the SYRKO CAD system description of Daimler-Benz AG for the modeling of external surfaces of the car body [6].

---

[1] from its German origin

Everything is designed with humans in mind; a product is tailored to the needs of its consumer. Products are created by means of material and spiritual production. They fill the person's living space—a place where he reveals the product's content and gives it an appropriate assessment. However, man evaluates the finished product by its necessity, in the aggregate of all its qualities. The usefulness of the product is primarily obtained through the visual communication channel. In this sense, the perceived shape of the product expresses, in its characteristics, the main qualitative indicators. As a rule, a harmonious form is an adequate reflection of the main technical, operational, and aesthetic qualities of an industrial product. In turn, the aesthetic qualities of the shape, embodied in design tools, play an important role in the product's sales. "Commercial success of any product depends largely on its design" [23]. Furthermore, a large part of product design can be attributed to its qualitative surface geometry. High aesthetic qualities of the shape contribute to the evaluation of the car by the buyer, and also become significant for many other industrial design objects. In the context of a car's industrial design, its basic aesthetic aspect is understood as the ability of a shape based on the high quality of the geometric relationships of the individual shape modeling areas among themselves, positively influencing the psycho-emotional sphere of its user, and forming the advertising aesthetic image necessary for the purchase of goods. Its content depends on the given compositional structure of the shape, various aspects of which are oriented towards providing a positive emotional perception of the geometry of the product surface by a person in different subject-spatial conditions with any kind of lighting. In this case, it is implied that, regardless of the nature and conditions of illumination (e.g., scattered, direct incident light, daytime, night, etc.), the steadily smooth motion of the light flare on the surface of the car's shape should be an indicator of the ideal smoothness of the connection of individual shape modeling areas of the product. The high culture of smoothness of the transition between individual surfaces creates a visual purity of the shape, which, at the level of the visual communication channel, is transformed into an aesthetic quality.

It is important to emphasize that the aesthetic properties of the shape of the design product, with the ideal conjugations of the shape modeling surfaces, remain constant in any subject-spatial situations, do not collapse depending on the perceived conditions and experience of the product's aesthetic image. This is due to the fact the emotional-psychological state of the consumer is transformed in a different object-spatial environment.

Thus, aesthetics become an important component of the product. Clearly, smoother curves and surfaces are more aesthetic [25]. Actually, the term a *smooth curve* carries an aesthetic component. In this article, the basic criteria for smoothness are considered as aesthetic criteria in the evaluation of the quality of shapes.

In work [3], the curves, the aesthetic qualities of which are an important component of the consumer properties of the product, are defined as aesthetic functional curves. To assess the quality of aesthetic functional curves, a multi-criteria approach is proposed. The authors of [3] believe that, in the evaluation of a product's design, priority should be given to its fulfillment of the smoothness criteria. Expert assessment from the standpoint of the laws of technical aesthetics is valid only after assessment for smoothness or in the absence of the possibility of such an analysis.

**2. FairCurveModeler Software Package**

To visualize class *F* curves and surfaces, the FairCurveModeler software package was developed, implemented in AutoCAD [7], Fusion360 [8], nanoCAD, COMPAS [9], Mathematica [4] applications, and the Web FairCurveModeler cloud product [10]. Translating a program to other CAD systems using their API presents a laborious task. Also, the functional expansion of the FairCurveModeler software package in all its implementations also proves laborious.

In [11], an idea is proposed about the possibility of using a universal software platform that is invariant with respect to the specifics of certain CAD systems. As such a software platform, the use of a combination of Microsoft Excel VBA, the FairCurveModeler software complex, and computer algebra systems is proposed. The software platform is designed to perform the following tasks:

- *For visualization of high-quality curves and surfaces for any CAD system*. In this instance, there is no need to transfer the functionality of FairCurveModeler to other CAD using their API. It is enough to write tools for exchanging geometric models between Microsoft Excel and CAD or use standard exchange via a DXF file.
- *For development of specialized applications (applied CAD) using the functionality of FairCurveModeler.* To do this, a set of open source programs that implement the interface between Microsoft Excel and FairCurveModeler is provided.

A necessary component of the software platform and, in general, CAD, are also systems of computational mathematics. The necessity and importance of the mathematical apparatus for CAD were well understood by Samuel Geisberg and Mike Payne, who introduced the Mathcad package into PTC products [12]. Computer algebra systems, together with the FairCurveModeler software, are used in the following ways:

- Calculation of the Hermite data, necessary for the approximation of the analytic curve by NURBS curve and input in the CAD of the analytical curve in the form of the NURBS template [13]. Hermite data is represented in the form of a table of coordinates of the points of the support polyline, the vectors of the first derivatives, the positive curvature values, the lengths of the segments of the curve between the support points and the unit curvature vectors
- Deep and comprehensive analysis of the NURBS curves constructed using the smoothness criteria
- Analysis of the NURBS curve for compliance with the engineering target (for example, for analyzing the geometric parameters of the cam profile for matching the specified macro parameters and the dynamic characteristics of the cam mechanism).

Thus, the software platform is an integrated system integrating Microsoft Excel VBA, FairCurveModeler, and computer algebra systems.

**Description of the Software Platform**

A universal software platform for visualizing class *F* curves and surfaces and developing applied CAD products with functional curves and surfaces comprises the following components:

- FairCurveModeler, a program complex for visualization of class *F* curves
- Microsoft Excel VBA
- Computer algebra systems, Mathcad, and Mathematica

The method for constructing a high-quality curve is comprised of the following steps:

- A virtual curve (*v-curve*) of class $C^5$ is constructed on two types of Hermite data of the form by a support polyline or tangent polyline.
- On the *v-curve*, a Hermite data is formed in the form of a table of coordinates of points, tangent vectors, and curvature values.
- NURBzS curve (rational Bezier spline curve) or NURBS curve (non-uniform rational *B*-spline of high even degree *m*, where *m* = {6, 8, 10}) is constructed by using Hermite data isogeometrically while preserving the quality of the *v-curve*.

The theoretical foundations of the program complex are described in [14-17]. The software package consists of two parts: the FairCurve.exe COM component and the interface part.

Computer algebra systems Mathcad and Mathematica are necessary for the analysis of NURBS curves, data preparation for the approximation of analytical curves in FairCurveModeler, and for solving computational problems in applied CAD-systems.

**Integration of the Software Platform with Fusion360**

The concept of using a universal software platform allows you to transfer the functionality of the FairCurveModeler software complex to any CAD system without special labor. For example, this concept is easily implemented for the perspective CAD Fusion360, an ambitious project of Autodesk. The functionality of Fusion360 provides all stages of product design. The Fusion360 is essentially an automated engineer workstation. Therefore, if we supplement the Fusion360 with the visualization functions of the class *F* curves, the Fusion360 will become a self-sufficient high-class CAD system capable of modeling functional and aesthetic curves as well as product surfaces, up to the outer contours of supercars.

The FairCurveModeler software complex is implemented in Fusion360 in [8]. The platform is used to extend the functionality of the application "FairCurveModeler app Fusion360" and for the possibility of developing an application CAD system in an integrated system software platform withFusion360. To ensure integration with the API (Application Programming Interface) Fusion360 developed a plugin that provides the exchange of geometric objects with Microsoft Excel. As part of the software platform, the Microsoft Excel `FCModeler + Fusion360.xlsm` book is presented, which provides the transfer of NURBS models to Fusion360.

The technique of working with the software platform for visualizing class *F* curves in CAD system Fusion360 is described below.

**1. Construction of a *v-curve* on a support polyline**

a) Build a polyline in Fusion360.

b) Call the plugin using the `plug_in_FairCurveModeler` button in the `ADD-INS` tab of the Model workspace.

c) Turn on the switch to `Points` in the `FAIRCURVEMODELER` dialog of the application in the `Exchange Options` area.

d) Select the vertices of the polyline in succession.

e) Click the `Exchange` button in the `FAIRCURVEMODELER` dialog box.

f) The text of the object model will appear in the `Exchange Box` text box. Place the cursor in the `Exchange Box` text box, select all the text (use the `End Shift`, and `Home` keys) and copy to the clipboard (`Ctrl` and `Insert`).

g) Go to the Microsoft Excel `FCModeler.xlsm` workbook on the `Polyline` page. Paste the contents of the clipboard into the sheet area, starting with the cell `Set_XYZ (C12)`.

h) In the Topology area, select the unclosed checkbox (a broken line is not closed). In the `Type GD` area, select `Basic (polyline)`.

i) To create the *v-curve* on the polyline, click `Create Curve`. The program will build the curve of the curvature curve function on the Graphics page (Figure 1) and form the NURBzS curve model on the NURBS page.

**Fig. 1.** The graph of its curvature curve on the Graphics page.

There are two ways to enter the NURBS model into the Fusion360 project:
− Inputting of the NURBS model into the project by a plug-in;
− Inputting the NURBS model in the project via a DXF-file.

**Entering the NURBS model into the project by a plug-in**) To transfer the curve to Fusion360, go to the Excel workbook `FCModel + Fusion360.xlsm` (the `apps Fusion360` folder) on the Fusion360 page.

Click on the `from NURBS to Fusion360` button (Figure 2). The program will copy the NURBS model to the clipboard.

**Fig. 2.** Microsoft Excel `FCModel + Fusion360.xlsm`, page Fusion360.

b) Go to Fusion360.

c) Insert the text in the `Exchange Box`.

d) The application displays the message `Press OK to create Object`.

e) Click `OK` in the message window. Click `OK` in the `FAIRCURVEMODELER` dialog box.

The Fusion360 APIs provide the means to input curve models in the Fusion360 NURBS format `transient` (intermediate). However, the API Fusion360 lacks a method of directly entering a curve from the `transient` format into the project. To enter a curve in the project, the curve in the plugin is interpolated, and an internal representation spline is generated at the interpolated points.

f) The application will display the message "Standard curve successfully created on points of a fair curve, number of points = ... "

g) Click `OK`. The curve is constructed.

h) Check the quality of the curve. Select the curve. Select `Toggle Curvature Display` from the context menu.

i) The program will construct a curvature graph over the curve.

Importing NURBS models in the `transient` format makes it possible to compute any parameter, but does not allow you to include in the project an exact model of the constructed curve. It is only possible to introduce the model into the project indirectly by approximating the internal representation with a spline. The main drawback of the spline of the internal representation is that it does not preserve the values of the boundary parameters of the original curve.

To construct non-closed curves with preservation of the quality of the *v-curve*, the following technique is proposed.

**Entering NURBS Model in the Project via DXF-File**

a) After plotting the curve in the Microsoft Excel workbook `FCModel.xlsm`, go to the NURBS page.

b) Click the `ViewCVT` button. The program interpolates the curve and constructs the curvature graphs. The program also generates a DXF file for the NURBS curve. The DXF file is saved in the folder `C://FairCurveModeler_TEMP/temp/with the name 'r_out_dxf.dxf'/`.

c) Next, go to Fusion360.

d) Call the `plug_in_FairCurveModeler` application.

e) Set the switch to `Curve from DXF`.

f) Clean the `Change Box`.

g) The program will display the message `Press OK to create Object`.

h) Click `OK` in the message window. Click `OK` in the `FAIRCURVEMODELER` dialog box.

i) The program will generate a curve from the DXF file.

j) Because of the difference in the units of measure in Fusion360 and in the DXF file, the curve will be 10 times smaller than the original.

k) `Select > Modify > Scale`.

l) Select the curve in the object tree.

m) Set the `Zoom` settings.

`Scale Type = Uniform`
`Scale Factor = 10`

n) The curve passes through the vertices of the original polygonal line. Importing a NURBS model through a DXF file preserves the geometry of the curve.

**2. Construction of a *v-curve* on a Tangent Polyline**

a) Build an arbitrary polyline in Fusion360.

b) Run the `plug_in_FairCurveModeler` plug-in from `Add-Ins`.

c) Turn the switch to `Points` in the `FAIRCURVEMODELER` dialog of the application.

d) Select the vertices of the polyline in succession.

e) Click the `Exchange` button in the `FAIRCURVEMODELER` dialog box.

f) The object model will appear in the Exchange Box text box. Place the cursor in the `Exchange Box` text box, select all the text (use the `End`, `Shift`, and `Home` keys) and copy to the clipboard (`Ctrl + Insert`).

g) Go to the Microsoft Excel `FCModeler.xlsm` workbook on the `Polyline` page.

h) Paste the contents of the clipboard into the sheet area, beginning with the cell `Set_XYZ (C12)`. In the

Topology area, turn on the `Closed switch (polyline closed)`. In the `Type GD` area, enable `Tangent (polyline tangent)`.

i) To create a *v-curve* on a polyline, click on `Create Curve`.

j) The program will build a curve and form the NURBzS curve model on the NURBS page.

k) To transfer the curve to Fusion360, go to the Excel Workbook `FCModel + Fusion360.xlsm` (the `apps Fusion360` folder) on the Fusion360 page. Click the `from NURBS to Fusion360` button. The program will copy the NURBS model to the clipboard.

l) Go to Fusion360.

m) Enter the text in the `Exchange Box`.

n) The application displays the message `Press OK to create Object`.

o) Click `OK` in the message window. Click `OK` in the `FAIRCURVEMODELER` dialog box.

p) NURBS from the `transient` format is converted into a spline of the internal representation.

q) The application will display the message "Standard curve successfully created on points of a fair curve, number of points = ..."

r) Click `OK` in the message. The curve is constructed (Figure 3).

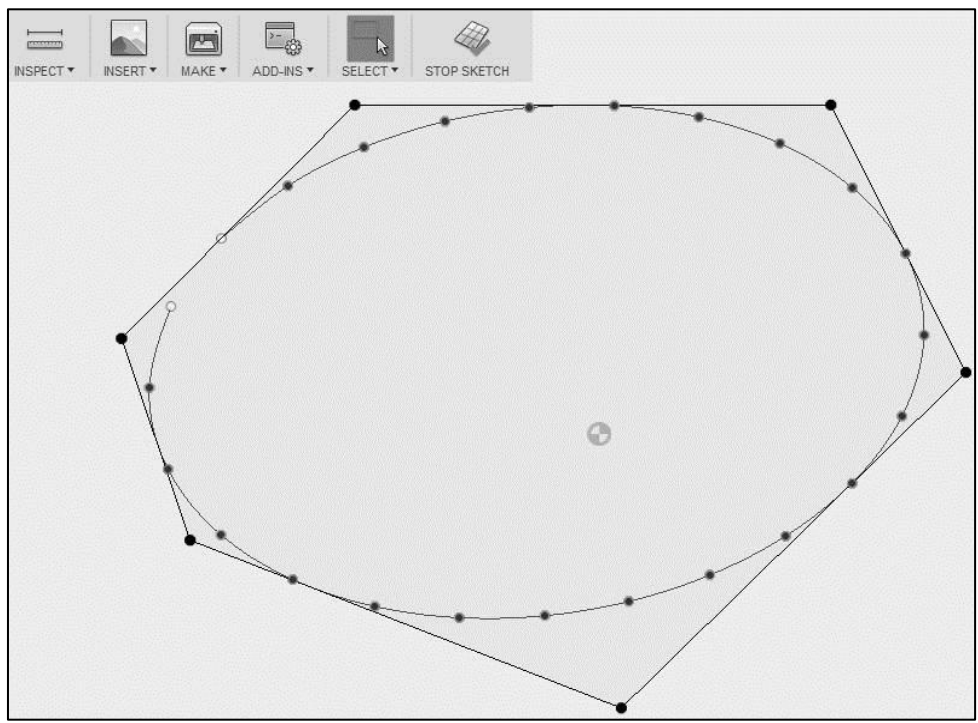

**Fig. 3.** The NURBS model approximated by the spline of the internal representation.

If the curve is closed, then when converting to an internal spline, the program generates an open curve. Close it by selecting the curve, then in the context menu select `Open / Close Spline Curve`. Check the quality of the curve. Select the curve and select `Toggle Curvature Display` from the context menu. Closed curves are converted while maintaining the quality of the *v-curve* (Figure 4).

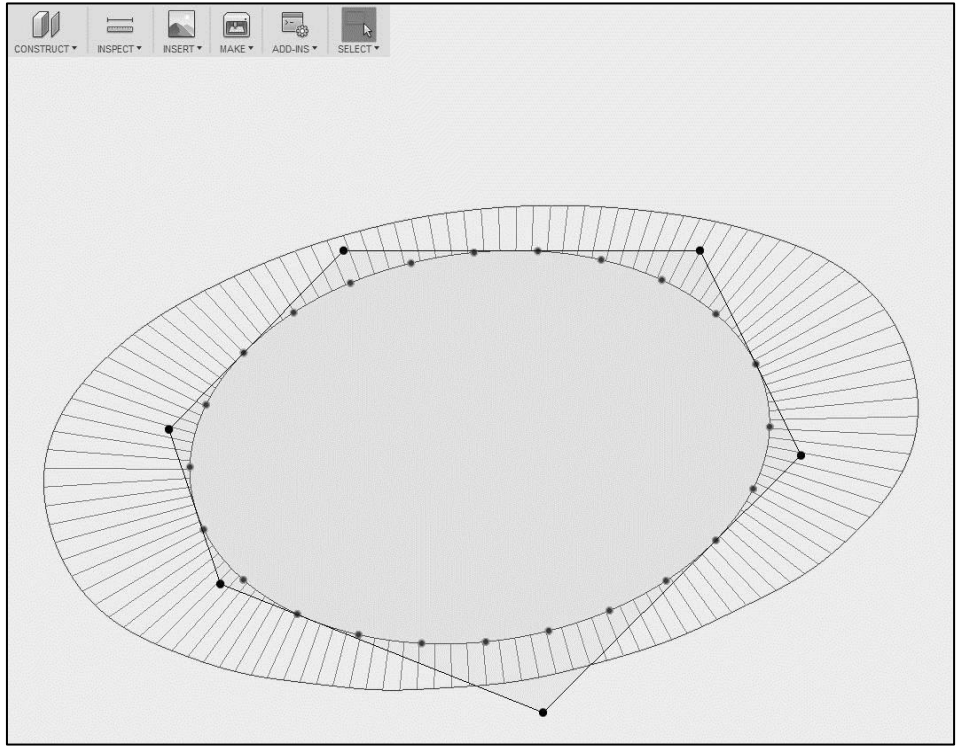

**Fig. 4.** Curve closure and quality control by plotting the curvature graph.

## 3. Construction of Aesthetic Curves

Obviously, class *F* curves are aesthetic unto themselves. However, along with just beautiful curves, there are "top models," log-aesthetic curves. Estimation of log-aesthetic curves is based on the mathematical characteristics of the forms revealed in real-world objects (e.g., butterfly wings) [18-19]. Aesthetic evaluation of curves is carried out by the laws of technical aesthetics and should reveal the aesthetic appropriateness of the curve in conjunction with the ability to meet the requirements of rationality [24]. In view of compositional order and structural coherence, the log-aesthetic curve has also expressive qualities. Hence, smoothness as a geometric property of the curve forms its expressive basis and visual purity. In terms of the emotional and psychological influences, smoothness contributes to the psychological comfort of a person, since it is associated with calmness, constancy, lack of aggression, as opposed to a broken line that gives rise to turbulent associations. To model aesthetic forms, the so-called log-aesthetic curves [20] are proposed, which have a linear curvature graph in the logarithmic scale. Several known spirals, including the clothoid, are particular cases of this class of curves. In [21], the broadest class of curves with a monotonic curvature function, superspirals, was proposed. The equations of these curves are expressed in terms of Gaussian hypergeometric functions and are numerically integrated by adaptive methods such as the Gauss-Kronrod method. The unique formula of the superspiral allows changing the three shape parameters *(a, b, c)* to get any of the known spirals and any spiral with a monotone change in curvature. In particular, for values of the parameters $a = 0.5$, $b = 1$, $c = 1$, the superspiral is a clothoid.

For importing analytical curves into CAD system, it is advisable to approximate them with NURBS curves. Thus, the NURBS template of the analytical curve is created. In the proposed software platform, Mathematica is used to calculate the most complicated superspiral formula. In Mathematica, a dynamic procedure for approximating and visualizing the superspiral was developed (Fig. 5). In a dynamic procedure, a Hermite data is formed in the form of a table of coordinates of the points of the support polyline, vectors of the first derivatives, positive curvature values, lengths of segments of the curve between the support points, and unit curvature vectors. The dynamic procedure uses the `FairCurve.exe` component directly and performs not only the preparation of the Hermite data for the approximation of the analytic curve but also its approximation by NURBS and estimate of the accuracy of the approximation. With a proper choice of the approximation parameters and the use of the Golden Mean Technique (clipping of the end sections with the form perturbation), one can achieve a high accuracy of approximation.

Fig. 5 shows the superspiral, representing the clothoid for $a = 0.5$, $b = 1$, $c = 1$. The clothoid is approximated by the *B*-spline curve of the eighth degree with the approximation parameters *Number of Points* = 16, the initial value of the parameter $s_0 = -1$, the first increment of the parameter $h_{s0} = 0.1$, and the last increment step is $h_{sk} = 1$. To eliminate the shape perturbation on the final section of the *B*-spline curve, the last three segments are removed.

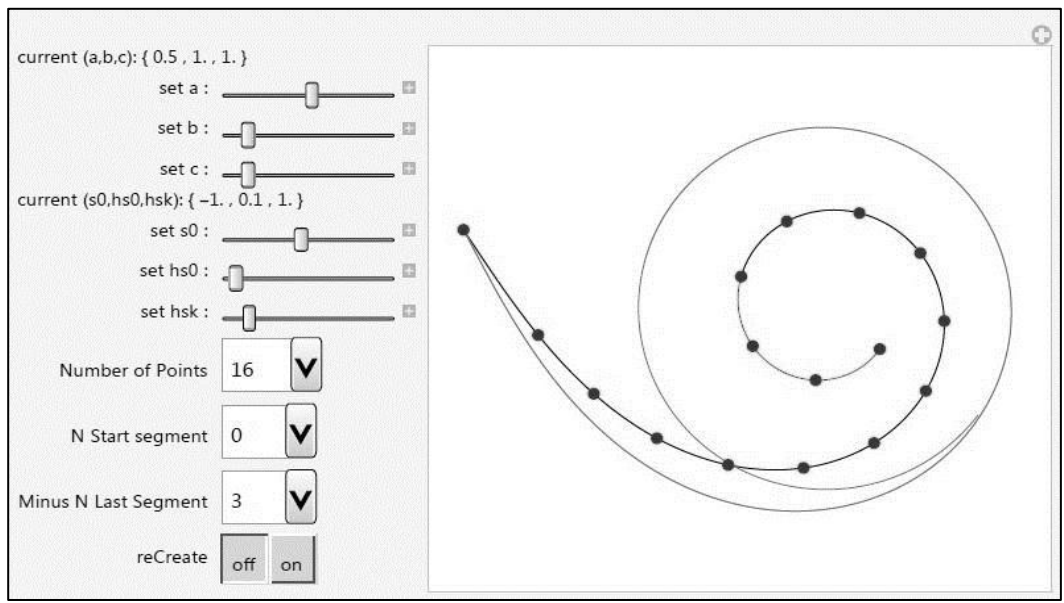

**Fig. 5.** Superspiral representing the clothoid. Approximation of clothoid by *B*-spline.

The dynamic procedure provides for the high quality and accuracy of NURBS-template construction. The *B*-spline curve visually coincides with the original clothoid. The curvatures of the *B*-spline curve and the original curve visually coincide as well. The program also gives a numerical estimate of the maximum deviations of the NURBS template from the original: max = max [0.00140453], min = min [-0.00417795]. The program generates the NURBS-model of the curve in Microsoft Excel format and exports it to an external file.

Next, to transfer the NURBS model from an external file to Fusion360, you must perform the following actions:

a) Open this file in Microsoft Excel, select the region with the model and copy it to the clipboard. Go to the `FCModeler.xlsm` book on the NURBS page. Paste the text from the clipboard, starting at cell `A2`.

b) Delete the last three spline segments. In the `Extract segments` area, set the parameters `Start segment = 0, Number of segments = 12`. Click the `Extract segments` button. Next, interpolate the curve to form the DXF file (`ViewCvt` button).

c) Go to Fusion360.

d) Call the application `plug_in_FairCurveModeler` in Excel.

e) Set the switch to `Curve from DXF`.

f) Clean the `Change Box`.

g) The program will display the message `Press OK to create Object`. Click `OK` in the message box. Click `OK` in the `FAIRCURVEMODELER` dialog box.

h) The program will generate a curve from the DXF file.

i) Due to the difference in the units of measurement in Fusion360 and in the DXF file, the curve will be 10 times smaller than the original.

j) Select `Modify-Scale`. Select a curve in the object tree. Set the following scaling parameters:
Scale Type = Uniform
Scale Factor = 10.

k) Superspiral will be displayed in the project. Check the quality of the superspiral by plotting the curve of its curvature (Fig. 6).

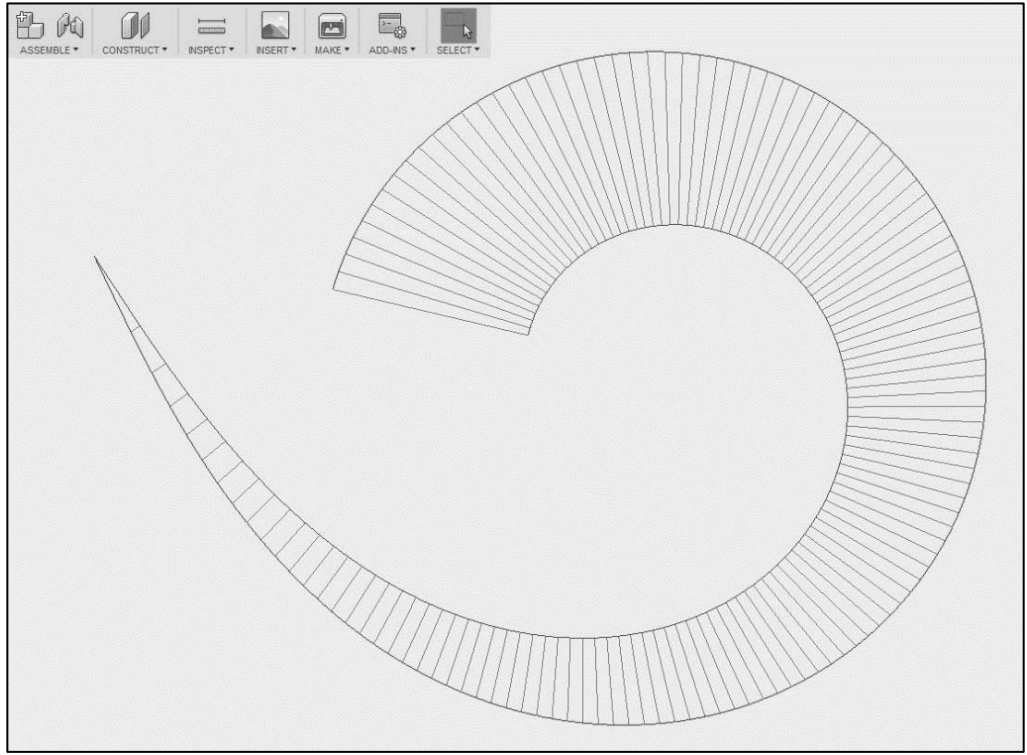

**Fig. 6**. Superspiral and its curvature function in the Fusion360 project.

## 4. Additional recommendations

Despite the self-sufficiency of Fusion360, it is recommended to use AutoCAD in addition to Fusion360 [22]. The authors of this article suggest that AutoCAD should be used to fix high-quality curves in three-dimensional models in order to avoid fitting the curves with a spline of the internal representation and preserve the high quality of class *F* curve. It is also recommended to use the application 'FairCurveModeler app AutoCAD' [7], which has the most complete functionality and allows modeling curves and surfaces. The application has an advanced toolkit for editing curves and surfaces for various types of data.

The software platform is complete with FairCurveModeler applications for CAD (Fusion360, AutoCAD, nanoCAD and COMPAS 3D). Before you start, you must transfer the activation code `Code_Activation.txt` from the application folder to the software platform folder `FairCurveModeler app Excel VBA`.

## 5. Conclusions

1. The program complex FairCurveModeler for visualization of high-quality curves and surfaces using criteria of smoothness is developed.
2. A universal software platform for visualizing class *F* curves and developing specialized applications for any CAD system based on the use of Microsoft Excel VBA, FairCurveModeler, and computer algebra systems is proposed.

3. The integration of the software platform with Fusion360 is demonstrated. The limitations of the Fusion360 API are revealed when the NURBS model is included in the project. The technique of transferring the exact NURBS model from the software platform to Fusion360 via a DXF file is demonstrated.

4. Application of the software platform for visualization of functional and log-aesthetic curves in integration with Fusion360 is demonstrated.

**Acknowledgment**

We should like to thank Rebecca Ramnauth of the Department of Computer Science at Long Island University and the Ravendesk Team, USA, who has generously given her valuable time to substantively edit and review this paper. Her care, competence, and conscientiousness are much appreciated. Additionally, we thank Ms. Elizabeth Jorgensen (USA) for useful remarks and suggestions.